\documentclass[conference]{IEEEtran}
\IEEEoverridecommandlockouts

\usepackage{booktabs} 
\usepackage{svg}
\usepackage{array} 
\usepackage{wrapfig} 
\usepackage{ragged2e} 
\usepackage{lipsum} 
\usepackage{cite}
\usepackage{amsmath,amssymb,amsfonts}
\usepackage{algorithmic}
\usepackage{graphicx}
\usepackage{textcomp}
\usepackage{xcolor}
\usepackage{todonotes}
\setuptodonotes{inline, color=blue!30}
\usepackage{caption}
\usepackage{siunitx}    
\begin{document}

\title{Non-programmers Assessing AI-Generated Code: A Case Study of Business Users Analyzing Data}

\author{
\IEEEauthorblockN{Yuvraj Virk, Dongyu Liu}
\IEEEauthorblockA{University of California, Davis, USA \\
ysvirk@ucdavis.edu, dyuliu@ucdavis.edu}
}

\maketitle

\begin{abstract}
Non-technical end-users increasingly rely on AI code generation to perform technical tasks like data analysis. However, large language models (LLMs) remain unreliable, and it is unclear whether end-users can effectively identify  model errors --- especially in realistic and domain-specific scenarios. We surveyed marketing and sales professionals to assess their ability to critically evaluate LLM-generated analyses of marketing data. Participants were shown natural language explanations of the AI’s code, repeatedly informed the AI often makes mistakes, and explicitly prompted to identify them. Yet, participants frequently failed to detect critical flaws that could compromise decision-making, many of which required no technical knowledge to recognize. To investigate why, we reformatted AI responses into clearly delineated steps and provided alternative approaches for each decision to support critical evaluation. While these changes had a positive effect, participants often struggled to reason through the AI's steps and alternatives. Our findings suggest that business professionals cannot reliably verify AI-generated data analyses on their own and explore reasons why to inform future designs. As non-programmers adopt code-generating AI for technical tasks, unreliable AI and insufficient human oversight poses risks of unsafe or low-quality decisions.

\end{abstract}


\maketitle
\section{Introduction}

Developers widely use LLMs for code generation. However, LLMs make mistakes that developers must recognize and correct. Increasingly, however, end-user \emph{non-programmers} also use AI-generated code, interacting through natural language or other intuitive methods.
In our study, 18 of 26 marketing/sales professionals reported using tools like ChatGPT to analyze data daily or weekly (which often involves code generation). Prior work explores code generation for non-programmers across domains such as UI generation \cite{liu2024crowdgenuienhancingllmbasedui,kolthoff2024zeroshotpromptingapproachesllmbased}, data analysis \cite{10.1145/3544548.3580817, ma-etal-2023-insightpilot, hong2024datainterpreterllmagent, guo2024dsagentautomateddatascience}, and robot programming \cite{Karli2024, yate2024}.

\begin{figure}
  \centering
  \includegraphics[scale=0.09]{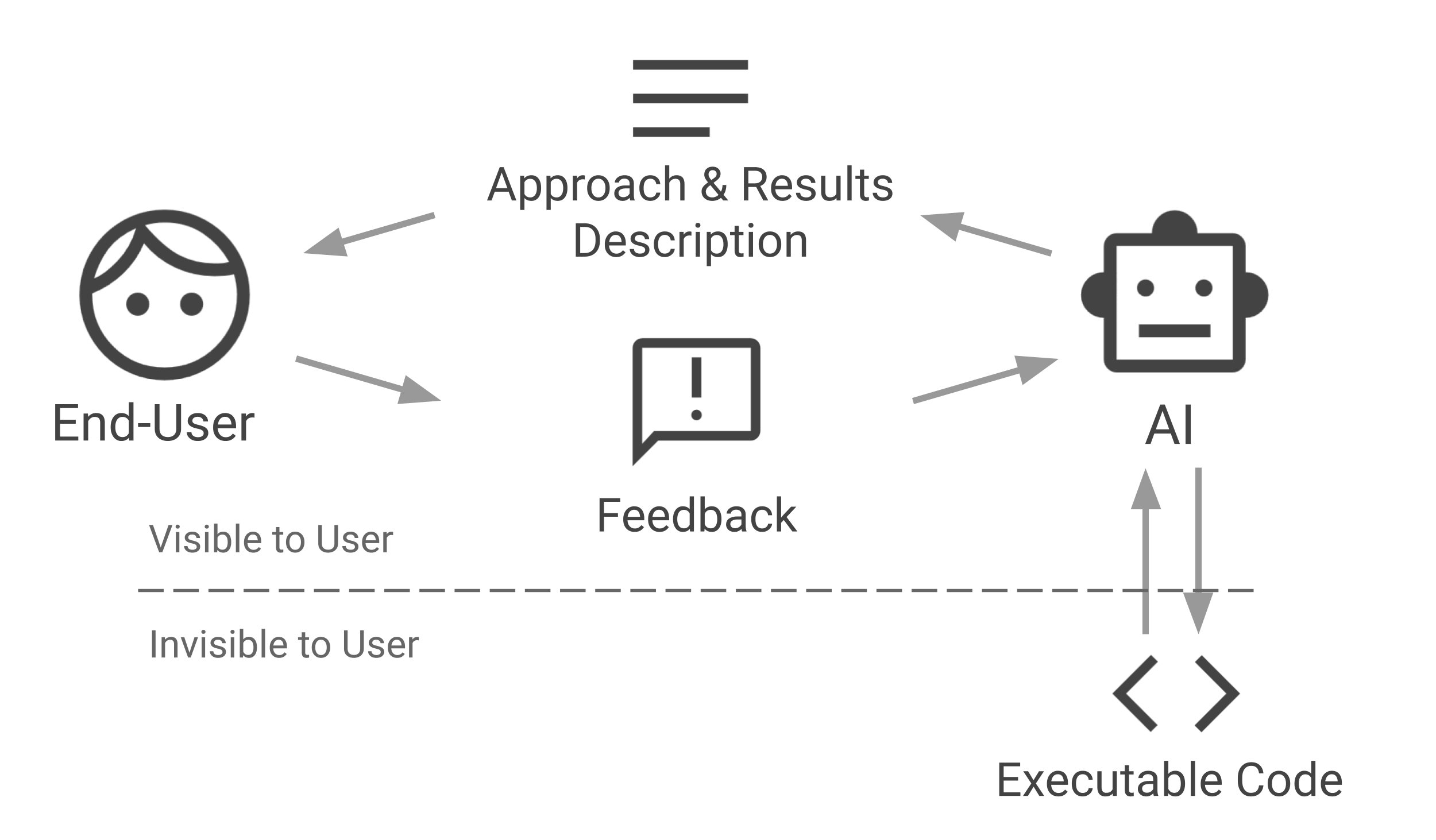}
  \caption{The end-user describes a task, the AI generates and runs code, then explains its approach and results. The end-user only views the natural language explanation.}
  \label{fig:model}
  \vspace{-1em}
\end{figure}

Yet, end-user non-programmers' current and potential code generation use is concerning: can they be reasonably expected to catch an LLM's mistakes \emph{without inspecting generated code}? Programs encode high-level approaches to achieve a programmer's goals; non-programmers might identify flaws in high-level approaches, but end-users we consider also lack relevant \emph{technical expertise} e.g. data science expertise for assessing data science techniques. Our study evaluates their ability to assess AI-generated code through natural language explanations of the AI's approach and results (Figure \ref{fig:model}). 

End-users we consider possess domain expertise and critical thinking ability, which they can use to assess an AI's technical output --- our study investigates to what extent and for realistic, domain-specific tasks. If they cannot reliably catch an AI's mistakes, and if the AI makes mistakes often enough, then our end-users \emph{can not safely complete technical tasks} under our setup. While AI code generation holds promise for end-users, it remains unclear how reliable AI must be to ensure safe use or whether they can complement its deficiencies.



To simulate a realistic use case, we focus on marketing and sales professionals analyzing data. These users commonly work with spreadsheets, but advanced analysis is outside their capabilities. Without access to dedicated analysts for day-to-day needs, business professionals resort to ad-hoc analyses \cite{gathani2022if}. While access to advanced analysis capabilities is important to business professionals, state-of-the-art LLMs remain unreliable for complex data analysis \cite{zhang-etal-2024-benchmarking-data, Lai2022DS1000}.

We first survey professionals to identify the types of flaws they detect in data analysis solutions. A second survey further explores why participants fail at assessing AI outputs. We hypothesize that AI decisions are not salient in outputs so restructured responses into clearly delineated steps. Second, we hypothesize AI decisions lack perceived alternatives so we present two plausible alternatives for each step.


\begin{figure*}
    \centering
    \includegraphics[width=0.9\linewidth]{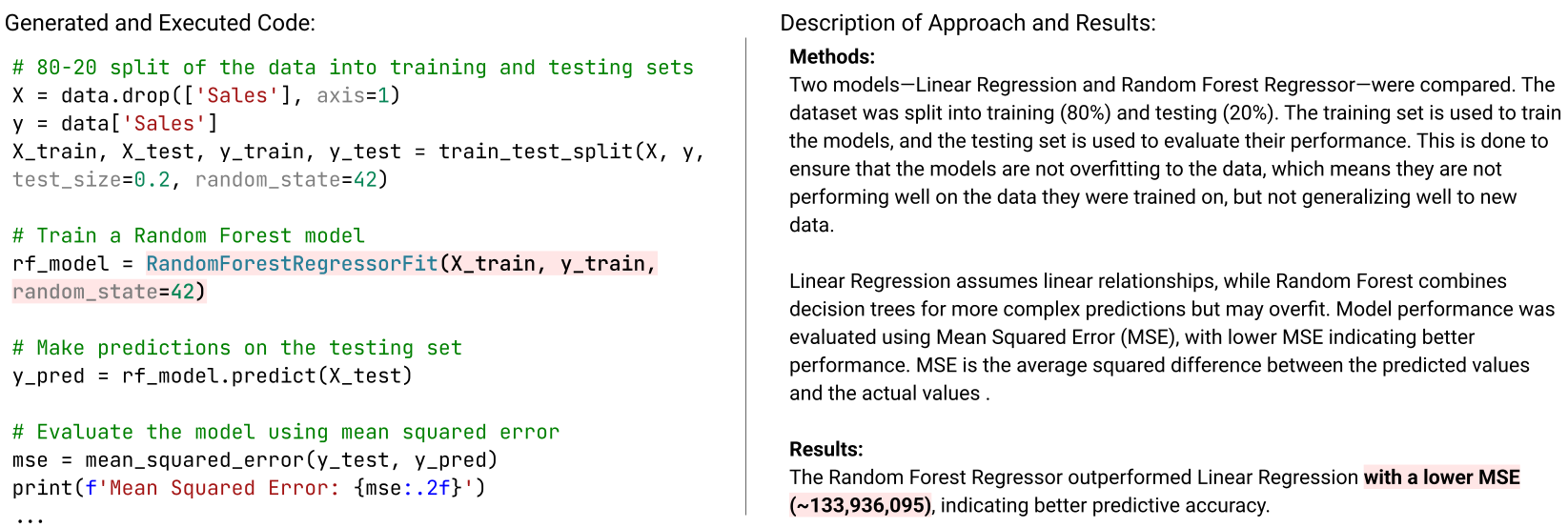}
    \caption{When prompted to select a model for predicting sales, the LLM generates and executes the code on the left. It generates the explanation on the right for participants. The highlighted code contains a hallucinated API name, a mistake non-programmers can not detect. The highlighted text on the right exhibits a flaw they can detect: MSE value is hard to interpret.}
    \label{fig:implementation_vs_approach}
    \vspace{-1em} 
\end{figure*}

In summary, our goal is understand whether business end-users can \emph{safely} use code generating AI for advanced data analysis and discover techniques to improve safety. Our study makes the following contributions:
\begin{enumerate}
    \item We conduct a survey of skeptical and motivated marketing/sales professionals (n=10) assessing 8 realistic, high-level data analysis tasks. We evaluate their ability to identify issues in the AI's response requiring domain expertise, technical expertise, or general critical thinking. We find they can identify valid flaws in each category but, overall, \emph{cannot be relied upon} to detect flaws in data analysis through our natural language explanations.
    \item Our second survey (n=18) evaluates the impact of explanation structure and presenting alternative steps on users’ ability to evaluate solution quality. These have a positive but inconsistent effect. Participants have a limited ability to assess alternatives and a low tolerance for deeply engaging with technical approaches.
\end{enumerate}





\section{Related Work}


\textbf{Studies of Programmers using Code LLMs}
Detecting errors in LLM-generated code is difficult for both novices and experts \cite{Vaithilingam2022, Dakhel2023}. Feldman and Anderson \cite{Feldman2024} studied non-expert programmers’ prompting and code-editing behaviors, completing the code evaluation step on their behalf. In contrast, we evaluate non-programmers' ability to judge natural language explanations of code.

Non-expert programmers can effectively use natural language representations to understand and fix AI-generated code, especially when explanations clarify jargon and reveal execution behavior \cite{ym2023pwrexploringrolerepresentations, yan2024intelliexplainenhancingconversationalcode}. Our study controls explanation quality (see Section~\ref{preliminary_study_setup}) to isolate non-programmer evaluation ability --- distinct from non-expert programmer abilities \cite{Feldman2024}.

Other work explores non-programmers using LLMs in domains like robot programming \cite{Karli2024} and spreadsheet programming \cite{Karsa2023}. Node-based visual editors and decomposing AI-generated code into helper and summary columns in spreadsheets enhance diagnosis abilities\cite{yate2024, Karsa2023}. Our findings align with this, suggesting that representations other than natural language may support more reliable evaluations.

Unlike these prior works that focus on low-level tasks, we study realistic high-level domain tasks. Our study isolates verification capability and controls for user over-reliance on AI outputs (see Section \ref{prelim_study_design}).

\noindent\textbf{AI Reliance}
\label{human-ai-collab-related-work}
Explanations, even when incorrect, increase reliance on AI recommendations \cite{10.1145/3411764.3445717}. To promote appropriate reliance, prior work proposes interventions such as verification reminders \cite{bo2024relyrelyevaluatinginterventions}. Rather than calibrating trust, we investigate whether non-technical users can detect flaws in technical outputs at all. We explicitly prime and incentivize users to distrust the AI. 

\begin{table*}[t]
\centering
\renewcommand{\arraystretch}{1.5} 
\setlength{\tabcolsep}{10pt} 
\caption{The table lists the task descriptions given to the LLM and participant, as well as an example of an author-identified shortcoming for the task. A flaw is \emph{safety critical} if it directly causes a poor or unsound decision when gone unnoticed.}
\resizebox{6.6in}{!}{
\begin{tabular}{@{}p{0.33\textwidth}p{0.3\textwidth}p{0.1\textwidth}p{0.1\textwidth}p{0.1\textwidth}p{0.1\textwidth}@{}}
\toprule
\textbf{Task} & \textbf{Example of Flaw in AI's Response} & \textbf{Safety Critical} & \textbf{Technical} & \textbf{Domain} & \textbf{Neither} \\
\midrule
1) Using the given data, add and test new variables that are important in helping predict sales. & Does not add variables for lagged effect of marketing efforts & X & \checkmark & \checkmark & X \\
2) Using the data with newly added factors, select a model that best predicts sales. & Difficult to interpret Mean Squared Error to evaluate model quality & \checkmark & \checkmark & X & \checkmark \\
3) Optimize drivers, within reasonable ranges, to maximize sales. Use the previous model. & Optimizes Competition's Online Impressions variable which is unrealistic & \checkmark & X & \checkmark & X \\
4) How much will the sales increase from average if you optimize for pricing within \$800-\$1200? Use the previous model. & Only tests sales outcome at \$1000 rather than entire range of pricing & \checkmark & \checkmark & X & \checkmark \\
5) Why do weeks with similar marketing or promotions show different sales outcomes? & Determines influential factors outside of dataset exist but does not discuss them e.g. holidays. & X & \checkmark & \checkmark & X \\
6) What factors drive sustained sales growth over multiple weeks, as opposed to only in the immediate week? & Measures correlation between sales and factors rather than between sales and lagged factors & \checkmark & \checkmark & X & X \\
7) Why are there sudden peaks in sales on certain dates? & Influence of holidays is unaccounted for & X & X & \checkmark & X \\
8) How do increasing promotions or discounts impact sales? & The linear model used does not account for the diminishing returns of marketing efforts & \checkmark & \checkmark & \checkmark & X \\
\bottomrule
\end{tabular}
}
\label{tab:tasks_shortcomings}
\vspace{-1em}
\end{table*}
\section{Study Setup}
\label{preliminary_study_setup}

This section details the scope of our natural language descriptions of code, how we generate them, and defines the categories of flaws participants are exposed to.

\textbf{Natural Language Descriptions.}
Participants see only natural language descriptions, with code hidden. Thus, code-only issues---like hallucinated API calls \cite{jain2024mitigatingcodellmhallucinations}---are out of scope. When the implementation is correct, the code reflects a high-level approach that can be evaluated in a comprehensive natural language description. Figure~\ref{fig:implementation_vs_approach} contrasts implementation and approach flaws. Thus, eliminating implementation errors is essential for non-programmer use. Our preliminary study examines approach-level flaws participants can detect.

\noindent\textbf{LLM-Generated Responses.}
We prompt \texttt{Llama-3.1-70B-Instruct-Turbo} in two phases:
\begin{enumerate}
\item \textit{Coding Phase.} The model iteratively generates code until it runs without execution errors \cite{yang2023intercode, ni2023leverlearningverifylanguagetocode}.
\item \textit{Explanation Phase.} The model explains the code and output in a “Methods” and “Results” format (Figure~\ref{fig:implementation_vs_approach}). All explanations are manually reviewed and, when necessary, adjusted by the authors to ensure they faithfully and comprehensively describe the generated code while remaining accessible to non-technical audiences.
\end{enumerate}

\noindent\textbf{Categories of Flaws.}
To analyze the types of flaws participants identified, we use the following categories. A flaw is safety-critical if, when undetected, it could lead to poor or unsound decisions (e.g., using a highly inaccurate model for sales projections). Each flaw is also labeled by content type and may fall into multiple categories:\begin{enumerate}
    \item \textit{Technical Content.} Issues with data science-related content, such as recognizing that a test set is too small.
    \item \textit{Domain Content.} Issues with marketing-related content, such as noticing the exclusion of holidays in an analysis of peak sales.
    \item \textit{Neither.} 
    Issues not necessarily tied to technical or domain knowledge, such as difficulty interpreting results e.g., unclear Mean Squared Error (MSE) values. These require general critical thinking.
\end{enumerate}
For example, both technical experience and critical thinking could be used to identify the ``unclear MSE values" flaw in Task 2, so it falls under ``Technical" and ``Neither."
\section{Preliminary Survey of End-User Evaluation Abilities}

We surveyed marketing and sales professionals to evaluate their ability to identify flaws in LLM-generated data analysis. Each task used a shared tabular dataset covering 27 weeks of marketing and sales data for a hypothetical TV company \cite{latentview_marketing_mix}. Participants received task prompts and LLM-generated solutions (Section~\ref{preliminary_study_setup}) and were asked to list any shortcomings, including explicit errors and omitted steps.

\subsection{Study Design}
\label{prelim_study_design}

\noindent\textbf{Tasks.} 
Due to limited programming expertise and a focus on end goals, end-user specifications can be high-level and underspecified \cite{10.1145/1922649.1922658, good-novice-underspecify}. Business professionals typically pose broad questions to derive actionable insights, which demand sophisticated analysis \cite{gathani2022if}. Gathani et al. \cite{gathani2022if} describe business users’ goals and workflows for data-driven marketing decisions. Based on these goals, we developed eight high-level data analysis tasks relevant to their needs. We decomposed the overarching task, “Optimize drivers to maximize sales,” into subtasks --- feature engineering, modeling, and optimization --- for easier evaluation. Each task exhibited at least one author-identified flaw in the LLM-generated response. Table \ref{tab:tasks_shortcomings} lists selected tasks and examples of their identified flaws.

\noindent\textbf{Priming, Incentives, and Engagement.} 
We took several steps to control for overconfidence and encourage thoroughness. Each task included a highlighted note stating: “The AI’s methods or results often have mistakes and can be improved on; some issues may be hard to notice.” Additionally, we offered a \$6 bonus (a 75\% increase over the \$8 base pay) to the top three performers. To ensure careful review, participants were first asked to restate the AI’s methods in their own words. 


\noindent\textbf{Participants.} 
We recruited 10 participants through Prolific~\cite{prolific_2024}, screening for: (1) no professional experience in programming or data science, (2) at least an undergraduate degree, (3) work in Marketing, Sales, or Business Development, (4) at least monthly use of marketing-related data to inform their business decisions, and (5) reside in the U.S. or U.K.

\subsection{Results}

\begin{figure} [htbp]
    \centering
    \includegraphics[width=0.97\linewidth]{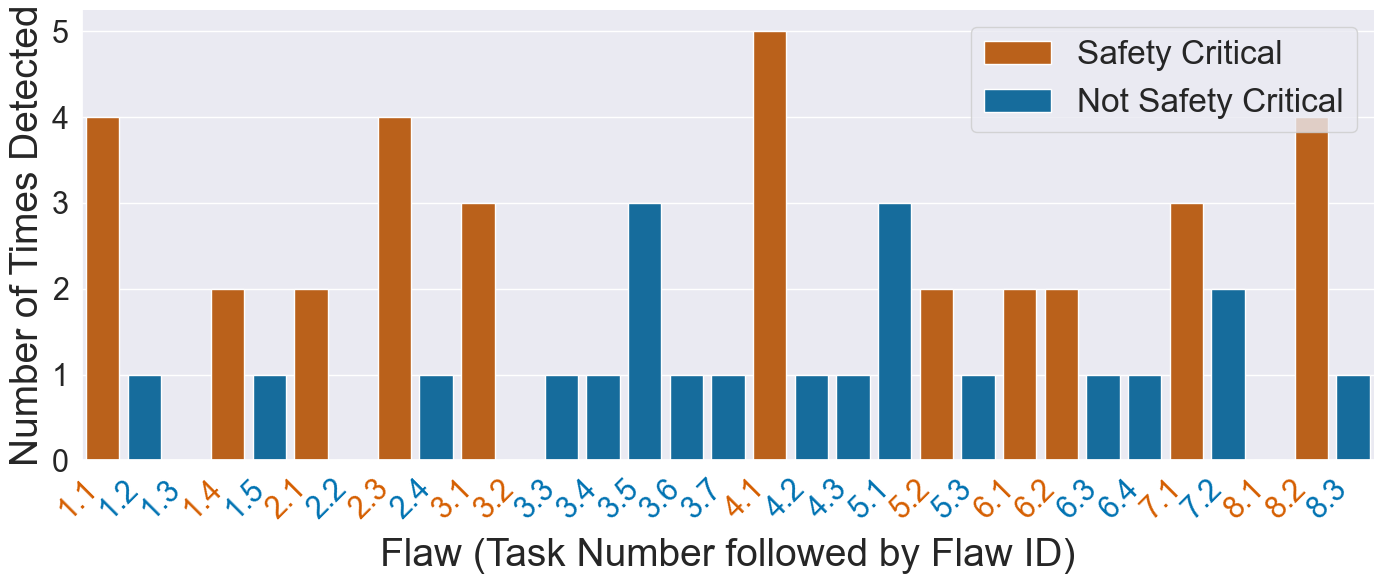}
    \caption{Total participants who identified each flaw. All flaws identified by authors and participants are labeled by their Task followed by a flaw ID e.g., 1.1 is Task 1, Flaw 1.}
    \label{fig:preliminary_bar_chart}
    \vspace{-1em}
\end{figure}


\noindent\emph{1) Quantitative Analysis:}\phantom{Invisible text line break}

\noindent\textbf{Participants struggled to identify safety-critical flaws.} Our analysis includes all flaws identified by participants and authors. Many were flagged by only one participant (Figure~\ref{fig:preliminary_bar_chart}), and none were consistently detected; some safety-critical issues were missed entirely by all participants.

\noindent\textbf{Participants demonstrated the ability to identify flaws involving technical information.} 
Of the 79 total flaws identified across all participants, 72.2\% involved technical content, 33.3\% involved domain-specific content, and 40.7\% could be identified without technical or domain expertise. Although detection was inconsistent, participants demonstrated the ability to critically assess technical information despite their limited technical backgrounds. The following qualitative analysis provides further insight into this capability.

\noindent\emph{2) Qualitative Analysis:}
We describe specific behaviors we observed in participant responses.
\begin{enumerate}
    \item \textbf{Broad vs. detailed discussion of methods}: Participants frequently discussed methods broadly rather than critically assessing detailed steps.
    \begin{quote}
        ``There doesn't seem to be any obvious deficiencies in this kind of methodology. Pitting two models against each other in two different kinds of datasets seemed to be a productive idea..." -- Marketing Associate evaluating Task 2 in Figure \ref{fig:implementation_vs_approach}
    \end{quote}
    \item \textbf{Failure to apply basic domain knowledge}: Participants sometimes failed to apply basic domain knowledge in the solution's technical context e.g., most missed that the AI ignored seasonality when analyzing peak sales (Flaw 7.1), despite its relevance in marketing.
    \item \textbf{Desire for concrete data}: Some participants wanted \emph{concrete data}, such as visualizations of forecasts generated by the AI's proposed model, instead of metrics like Mean Squared Error (MSE). They found abstract statistical techniques unconvincing.

    \begin{quote}
    ``There is no data or illustration to back up the results... we're just supposed to trust the results.'' -- Chief Marketing Officer (Task 2 in Figure \ref{fig:implementation_vs_approach})
    \end{quote}
    
    \item \textbf{Explicit vs. implicit shortcomings}: Participants more often flagged explicit flaws (e.g., incorrect logic) and overlooked missing elements (e.g., omitted features). For instance, no one identified features the AI should have included during feature engineering (Flaw 1.3).

    \item \textbf{Invalid critiques}: Some comments reflected misunderstandings of technical content in the AI's explanation.
    \begin{quote}
        ``testing information should be use in a higher \% than the training information” -- Real Estate Manager
    \end{quote}
    
    \item \textbf{Satisfaction with AI responses}: In several instances, participants found no issues and expressed satisfaction with the AI’s response.
\end{enumerate}

\section{Study on Effects of Formatting and Alternatives}
\begin{figure}[t]
  \centering
  \includegraphics[scale=0.11]{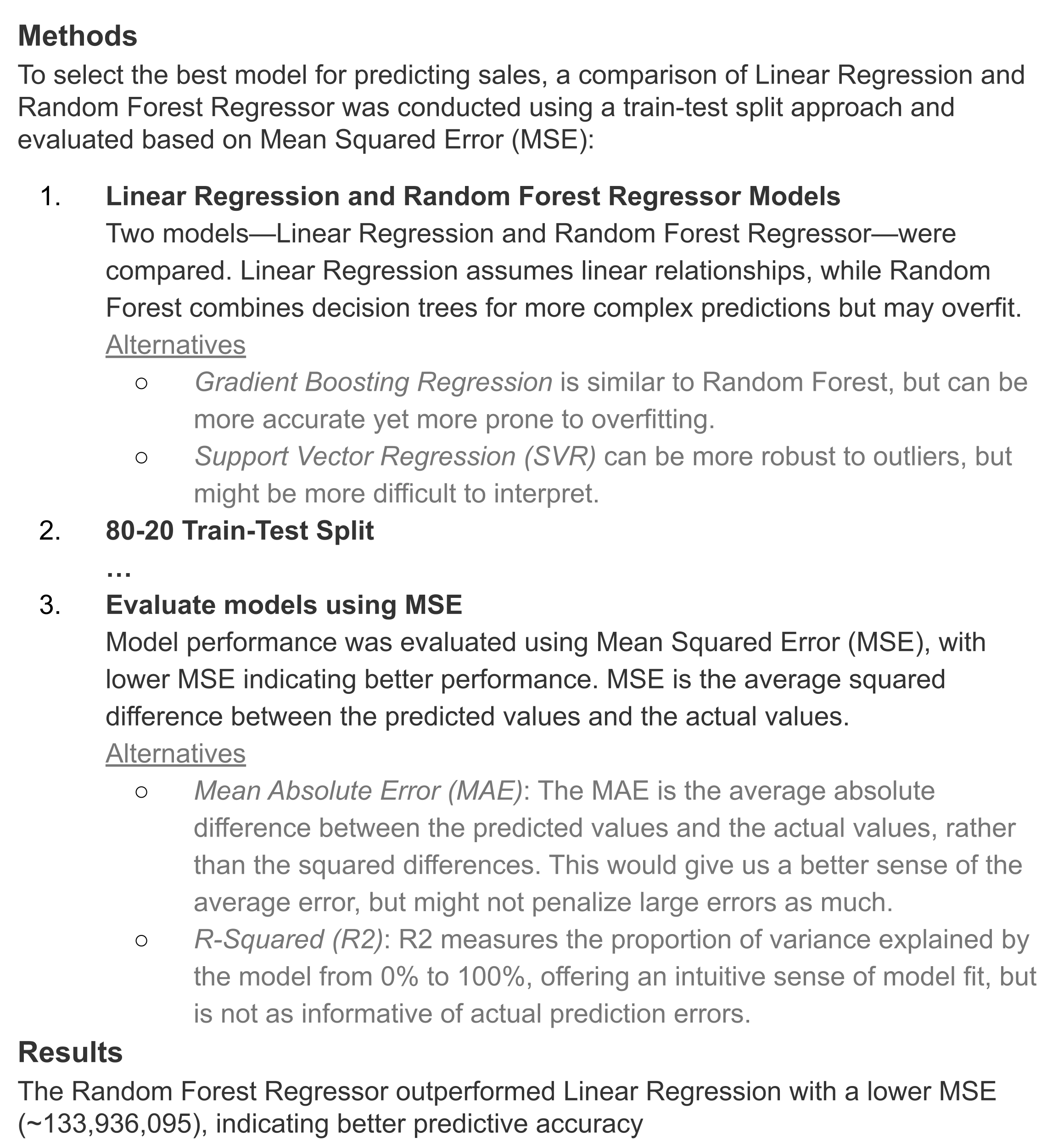}
  \caption{Enhancement of preliminary response in Figure \ref{fig:implementation_vs_approach}. It restructures the text to clearly delineate steps and presents two alternatives per step to support critical evaluation.}
  \label{fig:chunking_alts_example}
  \vspace{-1em}
\end{figure}

Our study revealed patterns in participants' evaluation abilities, but many reasons for overlooked flaws remain unclear. For example, in Task 3, the AI constrained sales optimization to historical driver ranges. However, increasing a driver (e.g., ad spend) beyond its historical maximum is an intuitive strategy. Detecting this issue requires critical thinking, not technical expertise, so why did all participants fail to recognize it? 

We propose two hypotheses. First, AI decisions are not \emph{salient}. End-users may miss AI decisions when buried in paragraph-style text. We restructured responses into clearly delineated steps using \texttt{Llama-3.1-70B-Instruct-Turbo}, followed by manual refinement. Second, AI decisions \emph{lack perceived alternatives}. Users may struggle to imagine alternative decisions, accepting the AI’s decisions without question. To address this, we presented two plausible alternatives for each decision, generated by the LLM and edited for clarity. Figure~\ref{fig:chunking_alts_example} shows an example combining both techniques.

\subsection{Study Design}
The second survey uses new AI-generated responses but otherwise mirrors the preliminary survey. We recruit 18 participants under the same criteria and use a within-subject design: each participant evaluates 4 formatted responses and 4 formatted responses with alternatives. To control for order effects, we apply a Latin Square design \cite{fisher1970statistical}, alternating conditions for each participant. Each condition-task pair is seen by 9 participants. To draw a balanced comparison, we randomly sample 9 participants from the preliminary study.
\subsection{Results}

\begin{figure}
    \centering
    \includegraphics[width=0.95\linewidth]{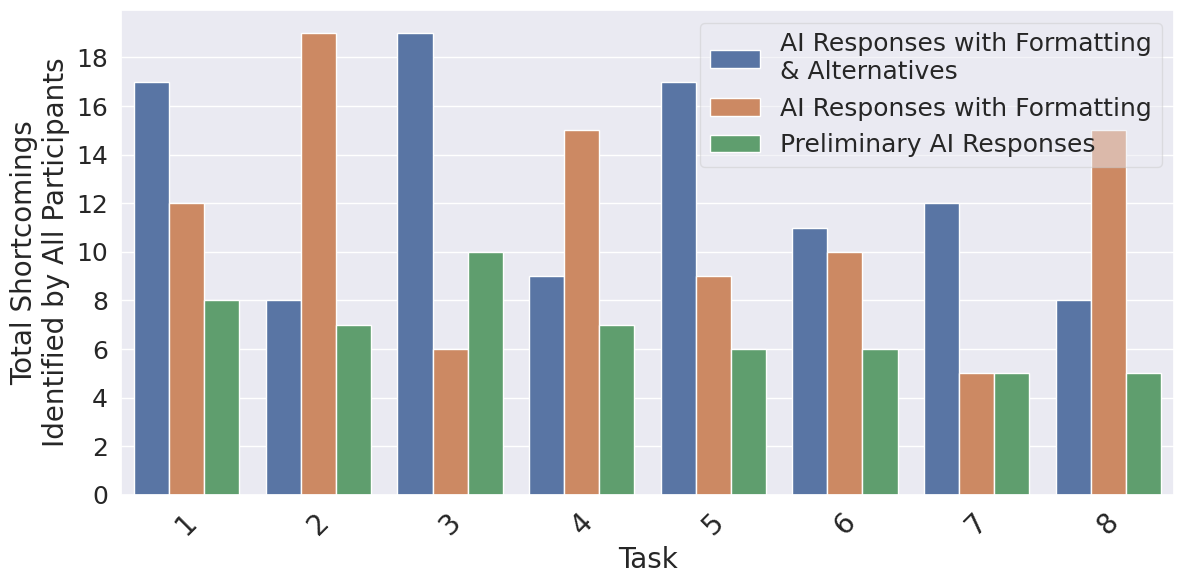}
    \caption{Total number of flaws identified by participants under different presentation strategies. The effect of formatted responses versus additionally displaying alternatives varies, but they are more often effective than the na\"ive response format.}
    \label{fig:AI_responses_comparision}
    \vspace{-0.25cm}
\end{figure}


\indent\textbf{Formatting helps notice simple flaws.} Participants produced more comprehensive descriptions of the AI's methods, often mirroring the AI's outline. However, formatting only substantially improved detection of Flaw 4.1, where the AI was tasked to optimize pricing over a range but evaluated only one price. With formatted responses, 8/9 participants identified the issue, versus 4/9 with formatting \& alternatives and 5/10 in the preliminary study. As this flaw was relatively obvious, simply breaking down the AI’s steps helped participants recognize it.

\textbf{Technical alternatives are hard to evaluate.} In many cases, alternatives had little impact, even when they directly pointed to shortcomings. Participants identified more issues in formatted responses for Tasks 2, 4, and 8 (Figure~\ref{fig:AI_responses_comparision}), where alternatives were technical and potentially overwhelming, while actual flaws were simpler. Two participants noted in open feedback that alternatives were overly focused on methods and too technical:
\begin{quote}
``[Alternatives are] helpful, broadly, but currently the focus on explaining the methodology rather than giving managers actionable insights. Many decison-makers would like to just see summary points with recommendations and a far more visual set of reports." -- Marketing manager 
\end{quote}


\textbf{Alternatives may encourage critical thinking.} On Tasks 1, 3, 5, and 7, participants shown alternatives identified the most flaws by a large margin. These flaws were not directly related to the alternatives, which were often too technical to evaluate. Still, Figure~\ref{fig:AI_responses_comparision} suggests alternatives may indirectly prompt deeper engagement. Given the unclear mechanism and small sample size, this remains early evidence.

\textbf{Individual differences affect critical evaluation ability.} Some participants correctly recognized alternatives as improvements over the AI's decision; others did not. This variation was unrelated to factors like professional experience or time spent per task.


%

\section{Discussion}

Our preliminary survey revealed that business professionals struggle to reliably detect safety-critical flaws in AI-generated data analyses. This suggests that they will make unsound decisions if they rely on unreliable models. 
The challenge is not simply a lack of technical knowledge; many flaws required only domain knowledge, or neither. Nor can the failure be attributed to overconfidence: participants were warned that the AI frequently errs, were directly prompted to identify mistakes, and were monetarily incentivized to do so. 

Instead, the difficulty is rooted in applying domain expertise or critical thinking to unfamiliar technical contexts. Even when the AI's approach was described with clarified jargon, clearly delineated steps, and alternatives, users struggled to reason through them. Some responses suggested they lacked the interest to engage deeply with the problem-solving approach, preferring visual summaries focused on actionable insights. 

Despite these challenges, participants identified a diverse set of valid flaws. To improve their ability to detect such issues, AI-generated explanations must better align with how end-users conceptualize data. Many participants preferred justifications grounded in graphs and concrete results over abstract textual descriptions of statistical methods. Additionally, our results suggest that careful formatting and presentation of alternatives can help. Alternatives must strike a balance: accessible enough yet rich enough to provoke critical thinking. Finally, we should design for wide individual variation in verification capability. While our study did not formally assess user traits, some participants leveraged alternatives effectively, while others detected few issues.

\section{Limitations \& Future work}

Our study evaluated verification capabilities using natural language descriptions of AI-generated code. While alternative representations --- such as visual diagrams explaining operations \cite{Karsa2023, yate2024} --- may better align with non-technical users' mental models, we focused on natural language descriptions. Generating visual explanations, especially for complex tasks like optimization procedures (Task 3), remains non-trivial.

We also treated user capabilities as fixed. Future work could explore skill-building methods that, while increasing the effort required by users, may empower users to more effectively engage with AI-generated outputs \cite{Feldman2024}. 

Our surveys are limited by their small sample sizes, especially in the second survey, which limit our ability to draw definitive conclusions about the effects of formatting and alternatives on verification. Additionally, findings may not generalize to other non-technical user groups with different skills or to tasks with different demands. Finally, participants might detect more errors in their usual work environments or specialized sub-domains of marketing.
\section{Conclusion}

Non-technical professionals are increasingly turning to code-generating AI for technical tasks like data analysis. However, our study finds that skeptical business professionals struggle to critically evaluate AI-generated data analyses, even when flaws are non-technical. While structured explanations and alternatives can help users notice and critically think through AI decisions, participants still face barriers in interest and ability to engage deeply in technical contexts.

Our findings suggest two complementary needs: (1) Developing highly reliable AI to reduce the burden on end-user non-programmers to verify outputs, and (2) Designing explanations that enhance end-user non-programmers' critical engagement through visualizations, salient decisions, and accessible alternatives. Addressing both helps end-user non-programmers more safely use AI-generated code in real-world contexts.

\section{Acknowledgments}
This work was supported in part by the U.S. National Science Foundation under Grant No. IIS- 2427770.





\bibliographystyle{IEEEtran}


\end{document}